**Exploring the relations between net benefits of IT projects and CIOs' perception of quality of software development disciplines**

assoc. prof. dr. Damjan Vavpotič[a,*], prof. dr. Marko Robnik-Šikonja[b], assoc. prof. dr. Tomaž Hovelja[c]

University of Ljubljana, Faculty of Computer and Information Science, Večna pot 113, SI-1000 Ljubljana, Slovenia

* corresponding author

[a] E-mail address: damjan.vavpotic@fri.uni-lj.si, Tel: +386 41 401 602

[b] E-mail address: marko.robnik@fri.uni-lj.si

[c] E-mail address: tomaz.hovelja@fri.uni-lj.si

**Exploring the relations between net benefits of IT projects and CIOs' perception of quality of software development disciplines**

**Abstract**

Software development enterprises are under consistent pressure to improve their management techniques and development processes. These are comprised of several disciplines like requirements acquisition, design, coding, testing, etc. that must be continuously improved and individually tailored to suit specific software development project. This paper presents an evaluation approach that enables the enterprises to increase development process net benefits by improving disciplines' quality and increasing developers' satisfaction. Our approach builds on Kano's model of quality. Based on an empirical study of top 1000 enterprises from Slovenia we find that application of software development methodologies in individual development disciplines significantly relates to net benefits of IT projects. The results show that different types of Kano quality are present in individual disciplines. Enterprises should be cautious when altering must-be quality disciplines like testing or deployment as they can significantly disrupt the established routines, cause great dissatisfaction between developers and significantly reduce benefits. On the other hand, changing the attractive quality disciplines like requirements acquisition can notably increase developers' satisfaction and benefits but is less likely to disrupt the established routines.

# 1 Introduction

Software development enterprises are under increasing pressure to compete in the global market. For this reason, their software development processes need to facilitate development of complex software products for demanding customers. Under these conditions, the software development process has to produce and support software products that fit many consumers, while at the same time achieve economic and design sustainability (Zdravkovic et al. 2015). This challenges software development enterprises to select and adapt their software development processes to support such software development projects (Bass 2016).

Software development processes comprise several disciplines that can be individually adapted and tailored to suit specific software development project (Vavpotič & Bajec 2009; Vavpotič & Hovelja 2012), such as requirements acquisition, design, coding & integration, testing, deployment, IT project management, etc. Improving the quality of these disciplines importantly affects software development enterprises success (Hovelja et al. 2015). However, quality of software development methodologies (SDM) is typically considered a linear concept meaning that satisfaction is considered to be proportional to the level of performance: the higher the performance, the higher the enterprise's (user of the SDM) satisfaction, and the lower the performance, the lower the enterprise's satisfaction (Chen 2012). This one-dimensional quality theory (Chen & Chuang 2008) limits the understanding of the quality of the software development disciplines and consequentially the ability of the enterprises to improve them.

This paper presents an evaluation approach for software development processes that considers quality as a nonlinear concept. It is based on a popular Kano's 2-dimensional model of quality. With Kano's model (Kano et al. 1984) we can evaluate enterprise's satisfaction with software development process. Such evaluation improves software development enterprise's ability to identify software development disciplines with improvement potential and facilitates the development of improvement strategies.

The remainder of the paper is organized as follows. In Section 2, we present the background and related work. We first introduce the importance of quality assessment of software development process and present the Kano model in this context. Next, we present the ReliefF and OrdEval algorithms that are particularly suitable for attributes measured on an ordered scale as is the case with the data collected in this research. In Section 3, we introduce our data and analytical approach supporting the Kano model. The empirical evaluation is covered in Section 4. In Section 5, we discuss the implications of the results and we conclude the paper in Section 6.

# 2 Background and related work

According to the CHAOS report (Standish Group 2015), only 29% of software development projects are completed on-time and on-budget, with all functions as initially specified. Another 52% of projects are completed and operational but over-budget, over the time estimate, and offer fewer functions than originally specified, which can severely affect the quality of the developed software. The quality of software development is addressed by several process maturity reference frameworks, such as ISO/IEC 15504 (Loon 2007) and the Capability Maturity Model Integration (CMMI) (Kneuper 2009). These frameworks recognize that the software development process can have various degrees of maturity, where increased process maturity results in increased predictability in relation to quality levels (Clarke & O'Connor 2012). Furthermore, they follow the process management premise that the quality of the system or product is highly influenced by the quality of the process used to develop and maintain it. Thus, they define maturity levels that embody this premise (Forrester et al., 2011). When it comes to defining the software development process, the claim that "one size fits all" is, in fact, a myth (Boehm & Turner 2003) due to the rich variation in situational contexts (Clarke & O'Connor 2012). Consequentially, it is important that the quality assessment of the software development process is performed for the key parts such as process disciplines and not only for the process as a whole (Hovelja et al. 2015). Such view is supported by several studies in the field of software development processes (Karlsson & Agerfalk 2009; Ralyte et al. 2003; Vavpotič & Bajec 2009).

Since its introduction in the 1980s, Kano's two-dimensional model (Kano et al. 1984) has become one of the most popular two-dimensional, nonlinear, asymmetrical models for evaluation of quality, and is adopted in a range of industries. The Kano model was developed to categorize the quality attributes based on their ability to satisfy customers' needs. The five Kano categories are must-be quality, one-dimensional quality, attractive quality, indifferent quality, and reverse quality. The must-be quality attributes have a logistic function shaped relation between quality and satisfaction (Fig. 1). They cause significant dissatisfaction when expectations are not fulfilled, however they do not contribute to the satisfaction of customers if quality level is adequate. The one-dimensional quality attributes have a positive linear-like relation between quality and satisfaction. They cause dissatisfaction when expectations are not fulfilled and satisfaction when they are fulfilled. The attractive quality attributes have an exponential-like relation between quality and satisfaction. They do not cause dissatisfaction when quality score is low but they strongly contribute to the satisfaction of customers when their quality score is high. The indifferent quality attributes do not have any significant relationship between quality score and satisfaction. The reverse quality attributes have a negative linear-like relation between quality score and satisfaction. They cause satisfaction when quality score is low and dissatisfaction when quality score is high. The graphs in Fig. 1 show idealized attributes, in reality, we might encounter different thresholds for satisfaction and dissatisfaction for attractive and must-be attributes, as well as different coefficients of linear growth for one-dimensional attributes. In fact, different groups of users might have different perceptions of an attribute, e.g., some taking it as one-dimensional and others as a must-be attribute.

Fig. 1 shows the impact of the must-be attribute (green line), one-dimensional attribute (blue line), and attractive attribute (red line) on satisfaction according to the Kano model. We assume that the scale of attributes and satisfaction are 7-point Likert scales, so satisfaction score of four is considered neutral.

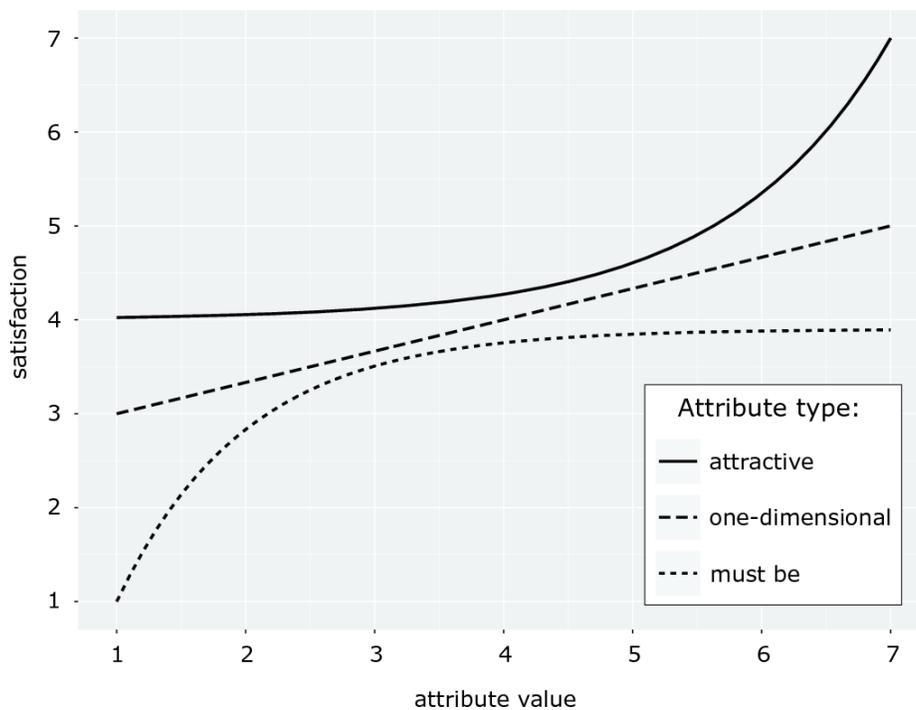

Figure 1: Graphs of idealized attribute types based on Kano quality categories.

Lofgren & Witell (2008) reviewed the research on the application of Kano's model over the last two decades, in which they found that 21 of the 28 studies used the original Kano questionnaire to classify attributes of quality. Kano's approach requires compilation of a questionnaire with a list of functional and dysfunctional questions for each attribute to observe the distribution of customer views. For example, customers are first asked how they would feel if a particular attribute were present or fulfilled (possible answers are a) satisfied, b) it should be that way, c) I am indifferent, d) I can live with it, or e) dissatisfied) and then how they would feel if that attribute were not present or unfulfilled (with the same possible answers). Response to both questions determines the nature of attribute according to Kano's model. Existing empirical studies have found this approach too complex and difficult to implement in real-world situations (Lofgren & Witell 2008).

Besides the original Kano questionnaire, several other approaches for classifying quality attributes were proposed (Chen 2012), such as penalty-reward contrast analysis (PRCA), importance grid analysis (IGA), direct classification method, and the moderated regression analysis. Practically most useful are regression methods such as PRCA, which introduce dummy variables to model nonlinear relationships between attributes and quality (Lin et al. 2010). These methods are theoretically unjustified and the resulting coefficients are not easy to comprehend, especially in the realistic circumstances with a presence of noise in questionnaires (Mikulic & Prebezac 2011). A practical advantage of the regression methods is that they directly analyze attribute-level customer satisfaction, which is much easier to collect in surveys than the list of functional and dysfunctional questions proposed by Kano.

In machine learning and data mining, the quality of attributes (also called features, independent variables, predictors, or input variables) is an important research question tackled in tasks such as attribute evaluation, attribute subset selection, and attribute ranking. Several measures have been proposed, which mostly estimate the quality of attributes through their predictive power concerning the response variable (also called dependent variable or class variable, in our case this is the satisfaction). Guyon & Elisseeff (2003) provide an overview of classical attribute selection approaches. Recently, the research in this area is focused on specialized measures, for example for specifics of bioinformatics (Bolon-Canedo et al. 2014) or big data (Zhao et al. 2017).

Simple attribute evaluation measures like Gini index (Breiman et al. 1984) take only dependence between one attribute and satisfaction into account. More advanced measures take into account conditional dependencies between a response variable and several attributes, the best known such measure being ReliefF (Robnik-Šikonja & Kononenko 2003). These attribute quality evaluation measures are concerned with the predictive power of attributes and, similarly to regression approaches, can detect important attributes. However, they do not take into account the specifics of ordered attributes (e.g., ordered attributes/questions in surveys) and cannot provide useful evaluation for each individual value of the attribute.

OrdEval (Robnik-Šikonja & Vanhoof 2007) is an attribute evaluation measure building upon ReliefF. It was initially developed for customer satisfaction research in marketing. The approach evaluates ordinal attributes, i.e. survey questions, based on their relation to the expected outcome. Different to ReliefF and other attribute evaluation measures it analyses each attribute's value separately and takes into account asymmetric effect an increase or decrease of attribute' value may have on the response. The approach allows a sort of what-if analysis. The algorithm computes conditional probabilities of the expected satisfaction upon changes in attribute values. For example, as an output, we get the conditional probability of higher satisfaction with a service if a satisfaction level of a certain attribute of that service would be higher. The approach also allows categorization of attributes according to the Kano model. Čufar et al. (2015) showed that using OrdEval one can construct simpler and substantially shorter questionnaire (one question per attribute) and still apply the Kano model. In our approach, we use ReliefF to identify important attributes and OrdEval to characterize them according to the Kano model, which allows us to analyze attribute-level customer satisfaction data.

Fig. 2 shows an illustration of three attributes according to Kano model: one-dimensional, must be, and attractive. Larger bars on the left-hand side of graphs shows downward reinforcements (e.g., the impact on satisfaction if attribute value changes from 7 to 6) and larger bars on the right-hand side shows upward reinforcements (e.g., the impact on satisfaction if attribute value changes from 6 to 7). The data underlying these graphs is almost ideal (uniform distribution of values and clear impact on the satisfaction), so we added some noise to show images that are more realistic. For the "one-dimensional" attribute all values show considerable reinforcements, both for increase and decrease of values. For "must be" attribute we can observe a strong jump in impact when the value changes between 1 and 2 as well as 2 and 3. For "attractive" attribute we notice a jump when the value changes from 6 to 7 or inverse. Note, that these are idealized attributes and in reality the effects may have different thresholds, they may be mixed within the same attribute (e.g., due to different perception of the same attribute by different subgroups of users), or may not be significant enough (e.g., due to low impact of attribute or insufficient number of users expressing certain score). To quantify the significance of the impact we added box-and-whiskers plots above each reinforcement bar which show the distribution of reinforcement scores under the condition that the impact of the attribute is random but with the same value distribution. The reinforcement bars stretching beyond whiskers (95% confidence interval estimated with bootstrapping) are therefore statistically significant in a sense that their effect is non-random with high probability. For example, due to added noise in "one-dimensional" attribute we see that the changes from 7 to 6 and 6 to 7 are not statistically significant. For real-world examples, the box-and-whiskers are typically wider than the ones shown for idealized attributes on Fig. 2.

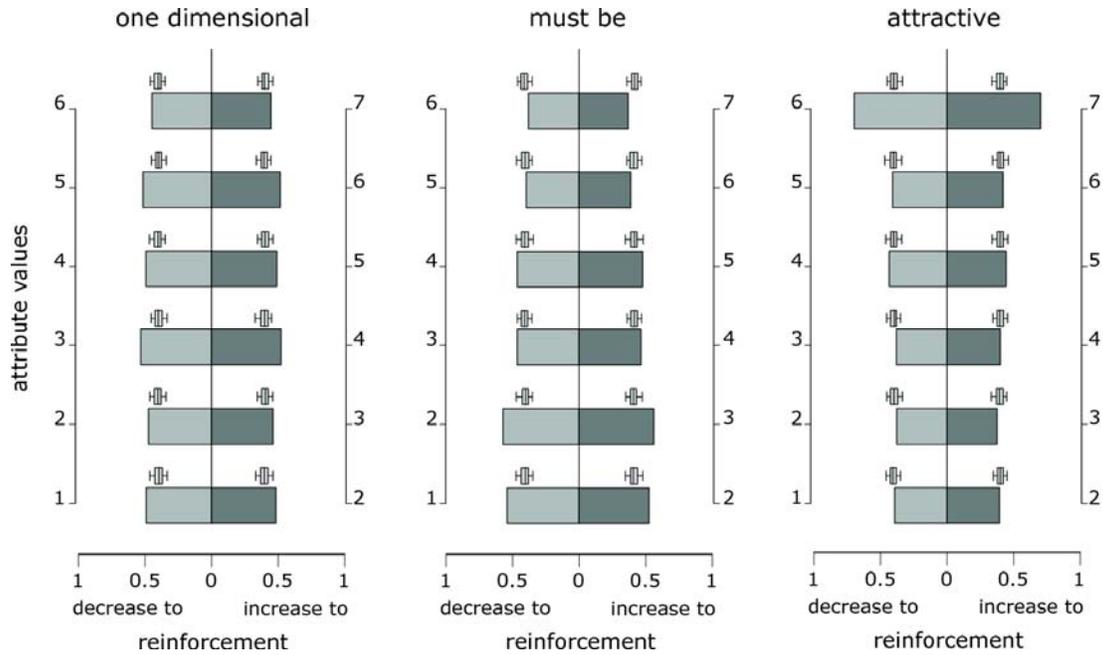

Figure 2: Visualization provided by the OrdEval algorithm for three idealized attributes: one-dimensional (left), must be (middle), and attractive (right).

The information the OrdEval algorithm provides has no parallel in standard approaches like multiple regression analysis. Firstly, there is substantial context sensitivity. Typically, the attributes are highly conditionally dependent upon the response and have to be evaluated in the context of other attributes. OrdEval is intrinsically contextual and assumes neither independence nor some fixed distribution of the attributes. Secondly, there is the ability to handle ordered attributes and ordered response and to use the information the ordering contains. The order of attribute's values contains information which is comparable but not the same as values of numerical attributes, e.g., values poor, good, very good, excellent are ordered in expressing certain attitude but this ordering is not necessarily linear. Thirdly, we have awareness of the meaning implied by the ordering of the answers and the positive (negative) correlation of changes between attribute values and the response (e.g., if the value of the attribute increases from poor to good, we have to be able to detect both positive and negative correlation to the change of the overall response value). Fourthly, OrdEval has the ability to handle each value of the attribute separately, e.g., for some attributes, the value of good and very good have an identical neutral impact on the response, value poor may have a strong negative, and value excellent a highly positive impact. We are able to observe and quantify each attribute's values separately and thereby identify important thresholds. Next to that, visualization of the output allows experts to use it as a powerful exploratory data analysis tool, e.g., to identify the type of attributes in Kano model. Also, the output is in the form of probabilities. Probability theory is commonly used and therefore the results in the form of probabilities are comprehensible and interpretable by a large audience and can also be used operationally. Finally, we have fast computation and robustness to noise and missing values.

## 3 Materials and methods

As the vast majority of other IT deployment studies in the literature (Bresnahan et al. 2002; Brynjolfsson et al. 2002; Hovelja et al. 2013; Mittal & Nault 2009) this study has also focused on the largest non-financial enterprises in a country. The starting population was the top 1000 enterprises in Slovenia based on the creation of added value in 2014 including IBM, Microsoft, Novartis, Goodyear, KPMG, etc. We sent the surveys to CIOs of studied enterprises and received 113 appropriately completed surveys. The survey was conducted from March until May 2016. Based on personal and phone communication with the CIOs involved in the study we found that the relatively low 11.3% response rate was mainly caused by the lack of time needed to fill out the questionnaire. Such response rate is typical for mail surveys conducted in enterprises in Slovenia (Hovelja 2008). The questionnaire asked participants for information about the characteristics and outcome of a recently completed important software project they had been involved in, regardless of its size and success. The key characteristics of our sample are as follows.

Fifty-eight percent of the projects had a budget of less than 100,000 EUR, 29% between 100,000 EUR and 500,000 EUR and 12% over 500,000 EUR while there was no response from the remaining 1%. On average 13.3 people were involved in the project of which 4.8 were external contractors. Fifteen percent of the projects lasted less than 3 months, 27% lasted between 3 and 6 months, 29% lasted between 6 months and a year, 21% lasted more than a year while there was no response from the remaining 8%. Twenty-eight percent of the deployed software products were custom solutions, 35% were customized local pre-packaged solutions, while the remaining 37% were customized pre-packaged solutions offered by international vendors. Eleven percent of the projects achieved all anticipated net benefits of the deployed solution, 64% partially achieved anticipated net benefits of the deployed solution, while the remaining 25% significantly failed to achieve most of the anticipated net benefits.

The key parts of the questionnaire focused on measuring the CIOs' satisfaction with the application of SDM in individual development disciplines and net benefits of IT projects. They were measured on a 7-point Likert scale ranging from 1 (strongly disagree) to 7 (strongly agree). The net benefits were defined in accordance with DeLone-Mclean model of IS success. The SDM was measured on the level of disciplines that were defined based on the well-established Rational unified process (Kruchten 2009) and included requirements acquisition, system design and architecture, coding and integration, testing, and deployment. The agreement statements about the net benefits of SDM disciplines for the deployed solution were analyzed and assigned to Kano quality categories.

We use two attribute evaluation measures: ReliefF and OrdEval, both implemented in R package CORElearn (Robnik-Šikonja & Savicky 2016). With ReliefF (Robnik-Šikonja & Kononenko 2003) we analyze the associations between CIOs' satisfaction with the application of SDM in individual development disciplines and net benefits of IT projects, taking into account possible attribute interdependencies. The ReliefF score near 0 or below is typical for disciplines having irrelevant net benefits, while positive values reveal disciplines with relevant associations with net benefits. The exact values of ReliefF score are problem dependent, therefore we avoid direct interpretation of numerical values and use ReliefF scores to rank the associations according to their importance.

With the OrdEval algorithm (Robnik-Šikonja & Vanhoof 2007) we analyze the impact of the level of attribute values i.e. CIOs' satisfaction scores for individual disciplines which allows us an inference about attribute characteristics according to Kano model. The visualization of OrdEval results can indicate the thresholds where attribute's values start having a strong positive or negative impact on the overall CIOs' satisfaction. The output of OrdEval are probabilities that an increase/decrease in the individual attribute's value will have an impact on the response variable. The intuition behind this approach is to approximate the mental decision process, taking place in each individual respondent, which forms a relationship between the attribute and the response. Namely, by statistically measuring a causal effect the change of an attribute's value has on the response value, we can perform probabilistic reasoning about the importance of the attribute's values, the type of the attribute, and determine which values are thresholds for a change of behavior. For each respondent, the OrdEval selects the most similar respondents and makes an inference based on the differences between them. For example, to evaluate the effect an increase of certain attribute value would have on the overall satisfaction, the algorithm computes the probability for such an effect from similar respondents with a larger value of that attribute. The overall process is repeated for a large enough number of respondents to get statistically valid results.

The methodology returns conditional probabilities called 'reinforcement factors'. These factors approximate the upward and downward reinforcement effect the particular attribute's value has on the satisfaction. For each value of the attribute, we obtain estimates of two conditional probabilities: the probability that the satisfaction value increases given the observed increase in the attribute's value (upward reinforcement), and the probability that the satisfaction value decreases given the observed decrease of the attribute's value (downward reinforcement). To take the context of other attributes into account, the probabilities are computed in a local context, from the most similar instances. The visualization of these factors with box-plots gives information about the role of each attribute, the importance of each value, and the threshold values. To understand the idea of the OrdEval algorithm, the attribute should not be treated as a whole. Rather, we shall observe the effect a single attribute's value may have.

For each reinforcement factor, the OrdEval computes confidence intervals which are plotted as box-and-whiskers plots above the obtained reinforcement factors. Reinforcement values outside the confidence intervals are statistically significant at 0.05 level. Missing values of survey questions are estimated from their class-conditional probabilities.

**4 Results**

In Fig. 3 we present the overall impact of each attribute on the overall satisfaction with the IT project taking into account possible attribute interdependencies (ReliefF).

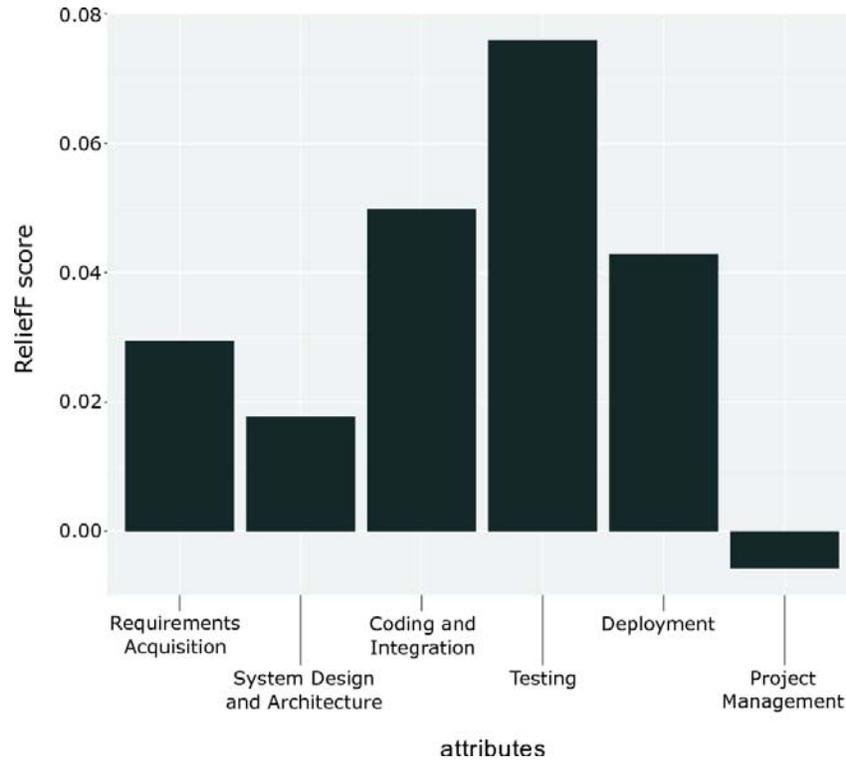

Figure 3: Association between CIOs' satisfaction with SDM application in disciplines and net benefits of IT projects (ReliefF)

The results show that application of SDM is importantly associated with the net benefits of IT projects in all disciplines (with the exception of project management). Based on our survey, *Testing* discipline has the strongest association with the net benefits of IT projects, while *Coding and integration* discipline is ranked second. Interestingly, CIOs satisfaction with the application of SDM in *Project management* discipline does not positively associate with net benefits of IT projects.

OrdEval method was used to classify SDM disciplines into five Kano quality categories. Figs. 4-9 show visualizations of OrdEval results for each discipline. When interpreting the results we focus primarily on statistically significant outcomes. Summarized results of how disciplines map to different Kano quality categories are provided in Table 1.

Table 1: OrdEval classification of SDM disciplines based on CIO's perceptions

| SDM discipline | Kano quality category |
| --- | --- |
| *Requirements acquisition* | Attractive quality and for a certain group of CIOs one-dimensional quality |
| *System design and architecture* | Inconclusive |
| *Coding and integration* | One-dimensional quality |
| *Testing* | Must-be quality and for a certain group of CIOs one-dimensional quality |
| *Deployment* | Must-be quality |
| *Project management* | Generally inconclusive and for a certain group of CIOs attractive quality |

Fig. 4 shows the statistically significant positive influence of SDM application in *Requirements acquisition* discipline on CIO satisfaction with the discipline when attribute value increases from 3 to 4, 4 to 5 and from 6 to 7 and statistically significant negative influence when the attribute value decreases from 4 to 3 and from 7 to 6. The top right and left bars show the strongest effects. This indicates that attractive quality of SDM application in *Requirements acquisition* discipline has a strong statistically significant influence on CIO satisfaction with the discipline, however, one-dimensional quality influence is also present.

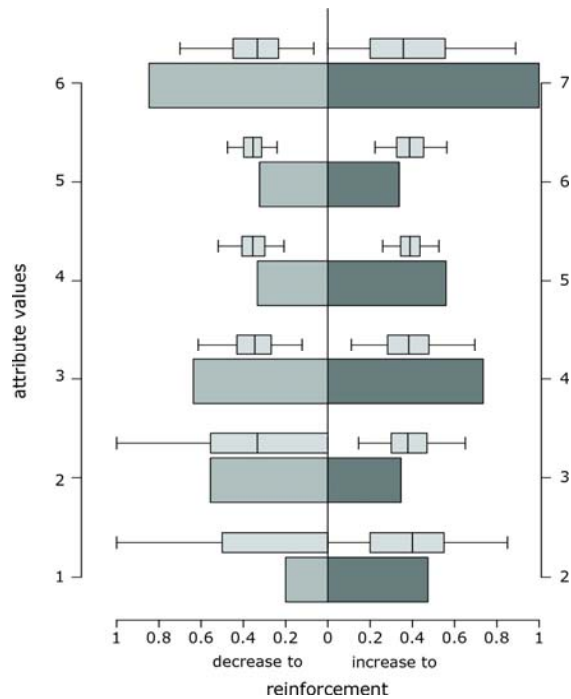

Figure 4: OrdEval results for *Requirements acquisition*

Fig. 5 shows that no statistically significant results were detected on the level of attribute values for *System design and architecture*. Thus we can't attribute any quality type to this discipline.

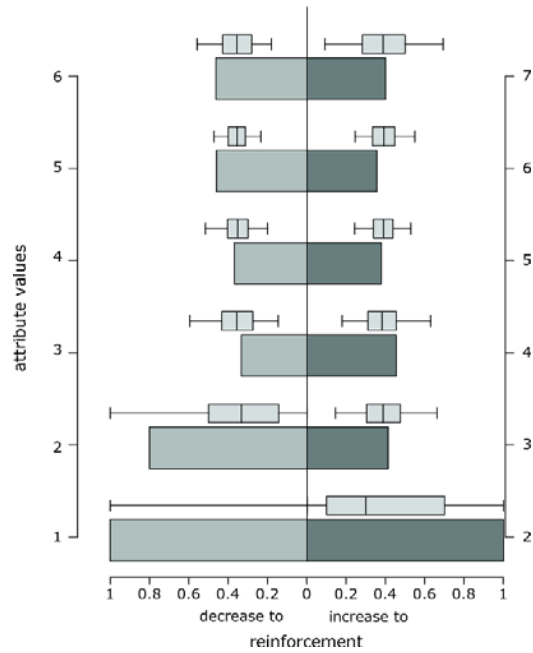

Figure 5: OrdEval results for *System design and architecture*

Fig. 6 shows the statistically significant positive influence of SDM application in *Coding and integration* discipline on CIO satisfaction with the discipline when attribute value increases from 2 to 3 and from 4 to 5 and statistically significant negative influence when the attribute value decreases from 5 to 4. This indicates a presence of the one-dimensional quality.

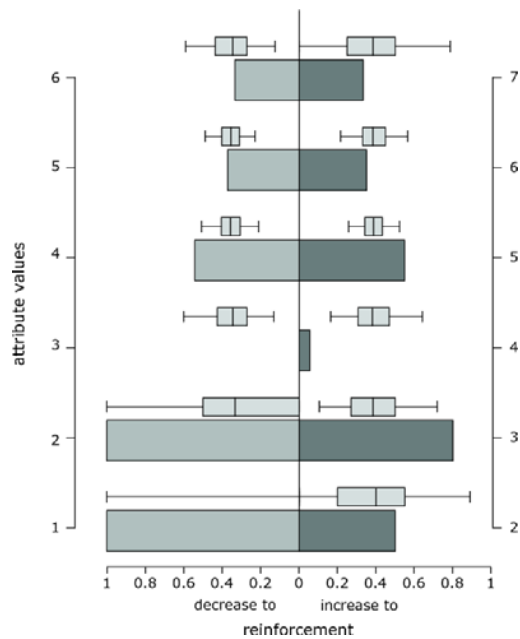

Figure 6: OrdEval results for *Coding and integration*

Fig. 7 shows the statistically significant influence of SDM application in *Testing* discipline on CIO satisfaction with the discipline when attribute value increases from 1 to 2 and from 5 to 6 or decreases from 6 to 5. The strongest effect shown in the bottom right bars indicates the must-be quality of SDM in *Testing*, while the effect shown by the bars in the second row from the top indicates that a certain group of CIOs perceive SDM in testing as one-dimensional quality.

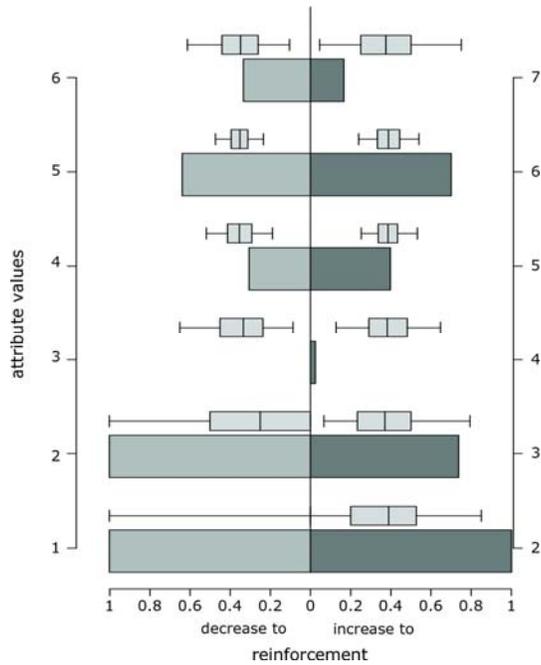

Figure 7: OrdEval results for *Testing*

SDM application in *Deployment* discipline is generally perceived as must-be quality as indicated by the bottom right bar showing statistically significant and strong effect on CIO satisfaction with the discipline in Fig. 8.

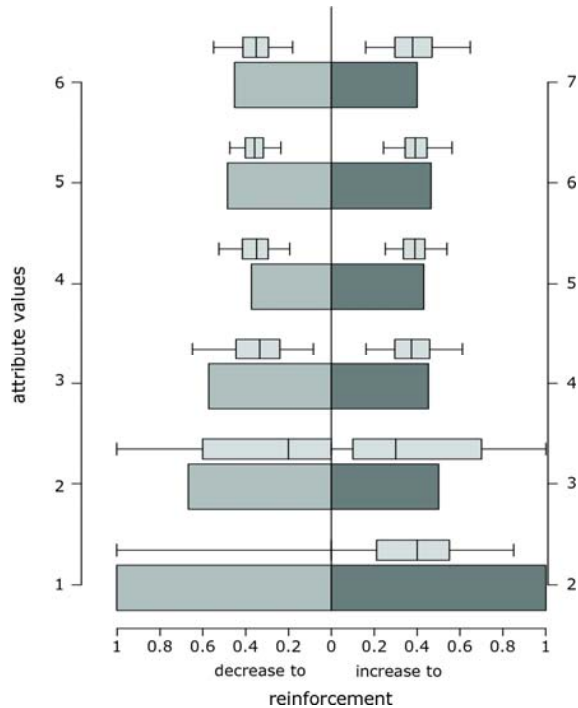

Figure 8: OrdEval results for *Deployment*

Fig. 9 shows the statistically significant positive influence of SDM application in *Project management* discipline on CIO satisfaction with the discipline when attribute value increases from 6 to 7. OrdEval results thus indicate a presence of attractive quality. Since RelieF score does not show that the impact of this attribute is relevant we can conclude that the benefits of *Project management* are on average inconclusive, but a certain group of CIOs perceives *Project management* as an attractive quality.

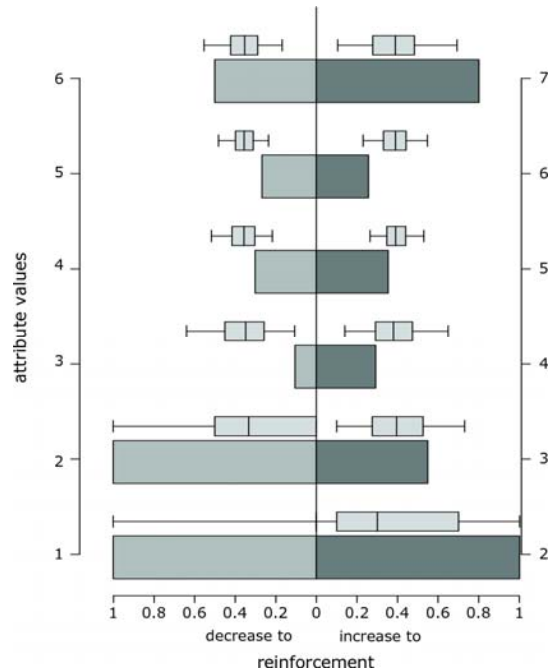

Figure 9: OrdEval results for *Project management*

## 5 Discussion

The attribute evaluation with ReliefF (Fig. 3) shows that application of SDM in *Requirements acquisition* discipline importantly affects the net benefits of IT projects. OrdEval further identifies the attractive Kano quality of SDM in *Requirements acquisition* discipline as having the highest statistically significant influence on CIO satisfaction (Fig. 4). These OrdEval results might be surprising since *Requirements acquisition* is considered one of the basic building blocks of the software development (Kruchten 2009; Maglyas et al. 2017) with high importance to software project success (Fernandez et al. 2017). The results can be explained by the fact that recently many new requirement acquisition techniques and approaches have been developed (Ernst et al. 2014; Lauesen & Kuhail 2012; Lucassen et al. 2016; Raspotnig & Opdahl 2013) and are probably perceived as innovations by the CIOs of the studied enterprises. Moreover, in an environment of ever changing customer requirements and technological changes there is a need for continuous reflection to decide on the best course of action (Kakar 2017). The newly developed approaches strive to improve communication between customers and development teams and try to gain a better common understanding of the problem domain, which importantly affects CIOs' satisfaction and IT projects net benefits.

The satisfaction with SDM application in *System design and architecture* discipline in terms of Kano quality varies considerably between CIOs. This is not surprising as enterprises use different SDM approaches in performing this discipline. Even when using similar SDM approaches different enterprises often focus on different categories of Kano quality. While formal modeling of software architecture can be a well-established and required approach for some enterprises (i.e. a must-be quality), it might be seen as an interesting but not required approach for other enterprises (i.e. an attractive quality). Consequentially, OrdEval detected no specific Kano quality (Figs. 3 and 4). Nevertheless, ReliefF (Fig. 3) clearly shows that CIO satisfaction with the discipline importantly positively associates with the net benefits of IT projects.

The ReliefF analysis shows that CIO satisfaction with the application of SDM in *Coding and integration* discipline importantly positively associates with the net benefits of IT projects. OrdEval further identifies one-dimensional Kano quality through the linear-like relation between the quality of SDM application in *Coding and integration* discipline and CIO's satisfaction with the discipline. *Coding and integration* discipline is largely defined by SDM approaches that systemize key processes importantly influencing net benefits of IT projects. For instance development lag that is a consequence of a continuous evolution of IS in large organizations (Neumann et al. 2014) can be considerably reduced by introducing the practice of continuous integration. Such approaches lead to higher level of efficiency in coding and integration discipline (Karvonen et al. 2017). This matches our findings regarding the positive contribution of SDM in *Coding and integration*.

The overall importance of *Testing* discipline is confirmed by ReliefF (Fig. 3), which shows that testing has the strongest positive association with net benefits of IT projects. OrdEval analysis of the SDM application in *Testing* shows two important Kano quality effects on CIO satisfaction with the discipline (Fig. 7). For one group of CIOs SDM application in *Testing* is considered must-be quality while another group of CIOs perceives it as a one-dimensional quality. We can conclude that basic levels of SDM application are required in *Testing* to enable project success, while higher levels of SDM application in *Testing* bring additional benefits. Such findings are consistent with findings of (Anand et al. 2013; Barr et al. 2015; Soetens et al. 2016) showing that additional benefits are related to SDM defining higher levels of testing automation.

As shown by ReliefF, the satisfaction of CIOs with the application of SDM in the *Deployment* discipline positively associates with net benefits of IT projects. These benefits probably stem from continuous deployment and may result in shorter time-to-market, continuous feedback, improved release reliability, increased customer satisfaction, and improved developer productivity (Rodriguez et al. 2017). OrdEval analysis of the SDM application in the *Deployment* discipline (Fig. 8) shows a must-be quality effect. We conclude that the SDM application to the *Deployment* discipline has to be established at least at the basic level since higher levels of SDM application in the *Deployment* discipline do not appear to be beneficial.

ReliefF (Fig. 3) shows that in general, the application of SDM in *Project management* does not importantly associate with net benefits of IT projects. However, OrdEval shows that a group of CIOs perceives application of SDM in *Project management* discipline as an attractive quality. Such findings are consistent with findings of (Wells 2012) who suggests that project management methodologies help inexperienced managers, who use them as a guidance tool, while they also help very experienced managers, who understand the value of promotion of standardization and uniformity of processes and procedures in multiple projects. However, according to (Wells 2012), the majority of project managers with medium experience do not perceive the project management methodology as helpful. Their perception of the benefits of a project management methodology is directly undermined by their tacit knowledge, intuitively steering project management decisions, and overriding the formal methodology directives.

## 6 Conclusion

The OrdEval and ReliefF methods proved useful for our analysis as they enabled a better understanding of Kano quality of individual SDM disciplines and their associations with net benefits of IT projects. According to our empirical study, the application of SDM in development processes disciplines, in general, has a strong impact on net benefits of IT projects with *Testing* providing the highest net benefits. Results show that enterprises should be especially cautious when altering *Testing* and *Deployment* i.e. must-be quality disciplines. Changes to these must-be quality disciplines can significantly disrupt the established routines and cause great dissatisfaction but are unlikely to significantly increase satisfaction. The opposite holds for *Requirements acquisition* i.e. an attractive-quality discipline where enterprises are encouraged to experiment. Recent innovations in the field can significantly increase the satisfaction but are less likely to cause dissatisfaction. Finally, *Coding and integration* discipline is considered to be a one-dimensional Kano quality. The management can expect that increasing its quality will result in a stable growth of benefits. The identification of these associations provides important new insights for better management of software development disciplines.

Our survey and analysis have certain limitations that should be considered when assessing the strength and generalization of the results. The included enterprises are not a random sample as the top thousand enterprises in Slovenia were surveyed. However, this group of enterprises presents a relevant study group due to its importance for the national economy. Larger datasets from multiple countries would improve the reliability of the results. Additionally, the respondents already knew the project outcome when they participated in our survey. This could have biased the responses especially if the projects were highly successful or highly unsuccessful. In line with similar studies (Jørgensen 2016), we tried to request mainly objective information about the project characteristics.

One of the avenues for improving our approach is to integrate it with the quantitative Kano models that have been recently developed. In order to address the subjective classification present in the Kano model, several quantitative

Kano models can be used such as Fuzzy Kano model, Continuous Fuzzy Kano model, Analytical Kano model etc. (Violante & Vezzetti 2017). Additionally, in future work, we intend to improve OrdEval algorithm to automatically merge values that do not have enough representatives for more reliable estimations of reinforcement factors. Another possible improvement that would increase the objectiveness of perception for different Kano qualities is to survey not only CIOs but also all other relevant stakeholders like project managers, developers and users.

**Acknowledgements**

The authors acknowledge the financial support from the Slovenian Research Agency (research core Funding No P6-0411 and No P2-0359) and from the European Union's Horizon 2020 research and innovation program under Grant Agreements No 825153 and 777204.

This is a pre-print of an article published in Business & Information Systems Engineering. The final authenticated version is available online at: https://doi.org/10.1007/s12599-019-00612-4